\documentclass[a4paper,twoside]{article}

\usepackage{epsfig}
\usepackage{subcaption}
\usepackage{calc}
\usepackage{amssymb}
\usepackage{amstext}
\usepackage{amsmath}
\usepackage{amsthm}
\usepackage{multicol}
\usepackage{pslatex}
\usepackage{apalike}
\usepackage{SCITEPRESS}     
\usepackage{color}
\usepackage{url}

\begin{document}

\title{Intelligent Luminaire based Real-time Indoor Positioning for Assisted Living}

\author{\authorname{Iuliana Marin\sup{1}, Maria Iuliana Bocicor\sup{1} and Arthur-Jozsef Molnar\sup{1}}
\affiliation{\sup{1}SC Info World SRL, Bucharest, Romania}
\email{\{iuliana.marin, iuliana.bocicor, arthur.molnar\}@infoworld.ro}
}

\keywords{Cyber-Physical System, Ambient Assisted Living, Received Signal Strength, Indoor Localisation, Trilateration, Levenberg-Marquardt Algorithm}

\abstract{This paper presents an experimental evaluation on the accuracy of indoor localisation. The research was carried out as part of a European Union project targeting the creation of ICT solutions for older adult care. Current expectation is that advances in technology will supplement the human workforce required for older adult care, improve their quality of life and decrease healthcare expenditure. The proposed approach is implemented in the form of a configurable cyber-physical system that enables indoor localization and monitoring of older adults living at home or in residential buildings. Hardware consists of custom developed luminaires with sensing, communication and processing capabilities. They replace the existing lighting infrastructure, do not look out of place and are cost effective. The luminaires record the strength of a Bluetooth signal emitted by a wearable device equipped by the monitored user. The system's software server uses trilateration to calculate the person's location based on known luminaire placement and recorded signal strengths. However, multipath fading caused by the presence of walls, furniture and other objects introduces localisation errors. Our previous experiments showed that room-level accuracy can be achieved using software-based filtering for a stationary subject. Our current objective is to assess system accuracy in the context of a moving subject, and ascertain whether room-level localization is feasible in real time.}

\onecolumn \maketitle \normalsize \setcounter{footnote}{0} \vfill

\section{\uppercase{Introduction}}
\label{sec:introduction}
In recent years, the advent of Internet of Things technologies have enabled the development of smarter solutions for various problems or aspects of life. Accurate indoor localisation and tracking of people proves to be one such aspect that is highly beneficial for multiple purposes \cite{zafari2019survey}. These include indoor navigation within large or complex structures such as shopping malls, airports, museums, offices or healthcare facilities; tracking staff or mobile equipment in factories or hospitals; facilitating cultural experiences augmented according to visitor context in exhibitions, museums or sports. One specific, but highly impactful use case for accurate indoor localisation is within the context of assisted living \cite{bianchi2018rssi}. 

The World Health Organisation expects the population of people over 60 to double worldwide by 2050 \cite{who15}. However, the same report finds that the latest advances in technology and medicine have yet to be adapted to address the problem, which is expected to impact healthcare and local government budgets. The study by \cite{meijer2013} shows a 4\% year over year increase in healthcare expenditure, from which an important part is expected to be allocated for the needs of the older population. This creates an opportunity where recent technological advances can be leveraged to improve the level of care and quality of life for older adults living within their own homes, as well as in nursing and residential facilities.

The present paper addresses the challenge of real-time indoor positioning in the context of the \textit{i-Light} \cite{marin2018} cyber-physical system for home monitoring and assisted living. The system was developed under funding from the European Union's Eurostars programme and uses intelligent luminaires that were developed as part of the project for indoor localisation. We present an experiment in which these luminaires are employed to obtain the real-time indoor position of a moving person. The experiments were performed in a three-room dwelling, with one luminaire deployed in each of the rooms. Luminaires incorporate a Bluetooth Low Energy (BLE) module used both for localisation and communication. 

Three scenarios were considered, in which the person moved from one room to another, in sequence, considering fixed start and finish positions within each room. Each scenario was repeated three times in the exact same conditions, for verification and refinement of experimental observations. Indoor localisation was obtained via trilateration in conjunction with nonlinear least squares optimisation. The intelligent luminaires collect the Received Signal Strength Index (RSSI) from the monitored user's smartphone every 5 seconds and transfer these to the system's software server, where the Levenberg-Marquardt algorithm \cite{Gavin2013TheLM} is used to determine the location of the monitored person.

Our initial indoor positioning experiments that detail several software-based approaches to improve localisation accuracy for stationary targets are presented in previous work \cite{marin2019indoor}. The main objective of our current work is to evaluate the system's performance with regard to indoor localisation of a moving subject. Accurate indoor positioning of moving subjects will lower the system's reaction times to potentially dangerous situations, and will facilitate the implementation of more advanced approaches for characterising user behaviour and detect changes indicative of cognitive impairment.

\section{\uppercase{Related Work}}
Solutions for indoor localisation and tracking have undergone great development in recent years, particularly due to their many and diverse applications. They are used in retail to improve customer experience, in manufacturing to increase efficiency by vehicle tracking or staff shifting, in culture and entertainment by improving visitor experience, as well as in healthcare for staff and device tracking, or visitor assistance. Although GPS technology has proven to be quite reliable outdoors, it does not represent a solution for indoor localisation due to poor signal strength \cite{ozsoy2013}. Obtaining accurate indoor localisation is a challenging task particularly because of the multitude of obstacles in the environment, as well as reflection and refraction that affect computations \cite{Miao2018}. This has led to significant progress in creating solutions for employing other types of technologies for indoors, such as Wi-Fi, Bluetooth, Global System for Mobile communications (GSM), Radio Frequency Identification, ultra wide band, acoustics and optics \cite{ta2017,lymberopoulos2015,xiao2016,Caron2017}. 

Techniques such as Time of Flight (ToF) and Time Difference of Arrival (TDoA) are popular when working with radio signals. ToF indicates the time a radio signal needs to travel from a transmitter to a receiver and back. The disadvantage of ToF is that the accuracy is limited by the precision of hardware timers \cite{Wibowo2009}. ToF can be improved by using the Bluetooth standard as a channel which mitigates the consequences of signal fading and inference \cite{giovanelli2018}.

TDoA uses the difference given by the arrival time of the signal sender and receiver \cite{Yu2019}. The difference is computed as being the multiplication between the speed of light in vacuum and the time difference. If there is an obstruction along the path of the signal, then this will result in wrong target positioning. Using the Angle of Arrival technique, the signal receiver determines the angle based on the direction of the incoming signal \cite{Wielandt2017}. Two or multiple nodes are set at known locations. The target position is given by the intersection of the lines given by the sensing nodes. This method is called triangulation. Infrared positioning is used by many electronic devices, but the accuracy is low, because rays cannot pass through walls or similar obstacles \cite{Randell2001}. Moreover, two infrared waves can interfere with each other. Ultrasound positioning is useful at night. The waves are characterised by low penetration through obstacles. As for infrared, a drawback is the interference of ultrasonic waves and reflection. A solution would be to combine it with ToF \cite{Qi2017}. 

Indoor positioning based on 3D cameras is useful and precise, but it does not offer privacy for the persons in the monitored area \cite{Li2010}. Intentional blurring using the bokeh effect was evaluated for determining indoor positioning based on the images taken with the user's mobile phone camera \cite{lee2019}. Blurring was proved to enhance position accuracy when determining the distance between a lamp with a single LED and the user's mobile phone. In this way, optical-power saturation is prevented and accuracy is enhanced. The coordinates of the person were obtained using the received signal strength and the angle of arrival. For the received signal strength, the optical power is determined using the area of the LED which appears in the photos. In the case of angle of arrival, the angles are determined using the received images, where the centre of the light coming from the LED was analysed. Using the two methods the 3D rectangular coordinates of the monitored person were obtained.

Both industrial players and the academic community have invested resources in solving the problem of accurate indoor positioning. One company that leverages the wide interest in indoor localisation and focuses on integrated location-based services is Navigine \cite{navigine}. They provide an indoor positioning platform allowing other developers and system integrators to build and create indoor navigation and tracking services. Their platform targets mobile developers, navigation software vendors, BLE beacon manufacturers, real time location systems solution providers, mapping companies, indoor navigation companies and others.

Sewio \cite{sewio} provides a complete platform for indoor tracking based on ultra-wideband technology. The system consists of a hardware component that includes signal transmitters called tags, which are attached to monitored targets, and signal receivers called anchors. Anchors are installed within the location and must cover the entire area for efficient tracking. The software solution of the platform includes various components for deployment and configuration, planning, optimal anchor distribution and type, real-time overview of the system's performance, remote settings updating, visualisation and analytic tools.

The indoor tracking solution offered by WI4B \cite{wi4b} consists of a mesh network of nodes deployed on location that are responsible with collecting and relaying data, managing performance and exchanging data with a host application. Wearable tags must be attached to the monitored persons or assets in order to allow them to be tracked. Another company that specialises in real time indoor positioning is Pointr \cite{pointr}, whose solutions range from indoor positioning and navigation to mapping, tracking and location based analytics. The positioning system employs beacons, inertial sensors and machine learning algorithms for both localisation and navigation. The real time localisation systems from infsoft \cite{infsoft} provide indoor navigation, geo-based assistance, analytics and tracking. These solutions use mainly Bluetooth beacons, Wi-Fi or ultra-wideband for indoor positioning.

Extensive academic research has also been conducted with regard to indoor localisation and various solutions have been proposed. Many employ RSSI \cite{luo2011,sadowski2018,maduskar2017}, which can be used to determine the distance between a target and the device measuring the received signal. However, one drawback of this method is that RSSI varies over time due to multipath fading \cite{pu2011}, degrading localisation accuracy. Several approaches to alleviate this problem have been proposed, such as the Simultaneous Localisation and Configuration algorithm proposed by \cite{bulten2016} as well as Kalman filter based approaches \cite{robesaat2017,sung2016}. 

\cite{trogh2015advanced} propose a system based on the Viterbi principle in conjunction with semantic data. They start from an RSSI fingerprinting technique, consisting of an offline and an online training phase and then employ the Viterbi algorithm to determine the most likely sequence of positions, using the sum of mean square errors between measurements and reference fingerprints. 

In contrast with the aforementioned studies and technologies, we present a system designed for home monitoring and assisted living, which includes an indoor localisation component based on intelligent luminaires. As opposed to other solutions for indoor positioning, our system does not require fingerprinting, tags or pre-installed anchors. The central components are a number of BLE-enabled devices that include sensing, communication and lighting modules \cite{marin2018}, and are deployed in the form of intelligent luminaires. 

The luminaires were created to replace existing light bulbs and to use existing energy infrastructure, thus permitting easy deployment, with no additional wires or devices, while offering reduced costs. The luminaires communicate via Wi-Fi with the system's server and collect localisation information from the monitored person's smartphone via the BLE wireless protocol. This information is relayed to the server in real-time, which uses trilateration and various optimisation algorithms to accurately compute the user's location.

\section{\uppercase{Real-time indoor localisation with intelligent luminaires}}
This section provides a concise description of the \textit{i-Light} cyber-physical system. The innovative aspects of the platform hardware and software are further detailed in our previous work \cite{iccp2017,marin2018,marin2019indoor}.

The platform hardware is based on luminaires that were custom developed as part of the project. They seamlessly replace regular light bulbs and use existing energy infrastructure. Each luminaire is either \textit{smart}, or \textit{dummy}, according to its processing and communication capabilities. Both types include a lighting module, so they can be used as regular light bulbs. Smart luminaires incorporate a sensing module for monitoring ambient conditions as well as a Wi-Fi and Bluetooth Low Energy communication module. Dummy luminaires are smaller and more cost-effective. Their main function is to scan the environment for Bluetooth signals and participate in the indoor localisation process by collecting RSSI measurements from the monitored person's device. Each dummy luminaire is connected to a smart luminaire, to which it sends gathered RSSI readings via BLE. Each smart luminaire may receive measurements from several dummy ones. Smart luminaires have the capability to communicate with the system's server via Wi-Fi and as such are used to process and relay aggregated measurements that include ambient conditions and localisation data to the software server for additional processing. The indoor positioning process takes place on the server, where trilateration and more advanced techniques, such as Kalman filters \cite{marin2019indoor} or optimisation algorithms are employed. In our previous work we have concluded that the accuracy required for the project's purposes can typically be achieved using up to three luminaires per room, depending on room size and shape. Each room should have one smart luminaire, to ensure the stability of the connection between smart and dummy luminaires, as well as that to the software server.

The platform includes a software server that provides a suite of custom developed, multi-platform applications. Together with the intelligent luminaire network, they enable the system to provide configurable home monitoring for older adults. The system uses ambient condition and RSSI data to provide indoor localisation, reports and advanced visualisations. Furthermore, the system provides notifications and real-time alerts to family members or caregivers in case a potential danger is detected. For example, this can happen when the monitored person is immobile, but not in their bed, or when ambient conditions such as temperature or volatile gas concentrations reach unhealthy levels.

\section{\uppercase{Experiments and results}}
The proposed experiment is the direct continuation of our previous work \cite{marin2019indoor}, where several software-based approaches were evaluated to improve the accuracy of indoor localisation for a stationary subject. We extend our evaluation to a moving subject, and study the impact of their movement on accuracy. Furthermore, we evaluate the consistency of recorded values with respect to errors by repeating each scenario three times.

Experiments were carried out within an apartment using \textit{n}-dimensional space trilateration, with nonlinear least squares optimisation. The algorithm's input is represented by the known positions of the luminaires and the distances from the monitored target to each of them. These distances are computed based on the recorded RSSI values, the path-loss exponent and the signal strength measured at a distance of one meter using the following lognormal model \cite{cherntanomwong2011indoor,huang2015,marin2020}:

\begin{equation}
\label{eq:localisation}
d=10^{\frac{A-RSSI}{10 \cdot n}}
\end{equation}

The significance of the variables in Formula \ref{eq:localisation} is as follows:
\begin{itemize}
    \item $n$ represents the path-loss exponent, ranging between 2 to 6 for indoor environments.
    \item $A$ is the signal strength expressed in dBm, measured at one meter. This parameter is experimentally computed once for each type of luminaire.
    \item $RSSI$ is the received signal strength index.
    \item $d$ is the computed distance. It is the distance between the luminaire that makes the measurement and the wearable Bluetooth device.
\end{itemize}

\begin{figure}[t]
    \centering
    \includegraphics[width=\linewidth]{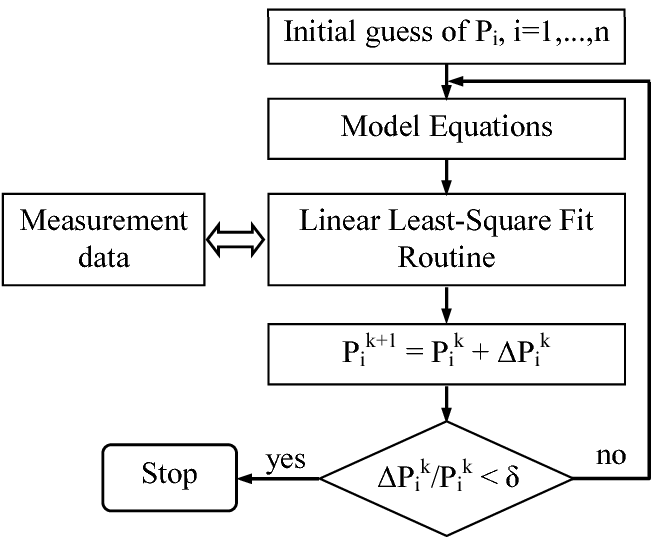}
    \caption{Optimized Levenberg-Marquadt algorithm flow \cite{LevenM2012}}
    \label{fig:LM}
\end{figure}

The output is the centroid that determines the position of the monitored person. At least three luminaires are required, along with their locations to allow the computation of the target person's position. The Levenberg-Marquardt algorithm is used to solve the nonlinear least squares problem \cite{Gavin2013TheLM}. It is a combination of the gradient descent and the Gauss-Newton methods for optimisation. Starting from an initial guess of the parameters, it optimises them in an iterative manner: when they are far from their optimal value, the algorithm behaves like gradient descent, while when they are closer, it switches to employing the Gauss-Newton method.

The parameters to be optimised are represented by the two dimensional coordinates of the monitored person's location. They can be determined from the following system of at least three\footnote{One equation per each luminaire.}  equations: $(x - x_{i})^2 + (y - y_{i})^2 = d_i^2$, where $(x_i, y_i)$ are the coordinates of the $i^{th}$ luminaire, $i \in \{1, 2, \cdots, n\}$, with $n \geq 3$ and $(x, y)$ is the (unknown) position of the monitored person. The objective function is thus a non-linear function of the parameter vector $P=[x, y]^T$ and the idea is to find the parameters such that a minimal error between the model and observed measurements is obtained.

The general form of the optimisation algorithm is described in Figure \ref{fig:LM}. It starts with an initial guess for the fitting parameter values ($P$). In each iteration the parameter vector is replaced by a new estimate ($P + \Delta$), according to the Levenberg-Marquardt formulae \cite{LevenM2012}. This process continues until the calculated step falls below a predefined limit.

\begin{figure}[t]
    \centering
    \includegraphics[width=\linewidth]{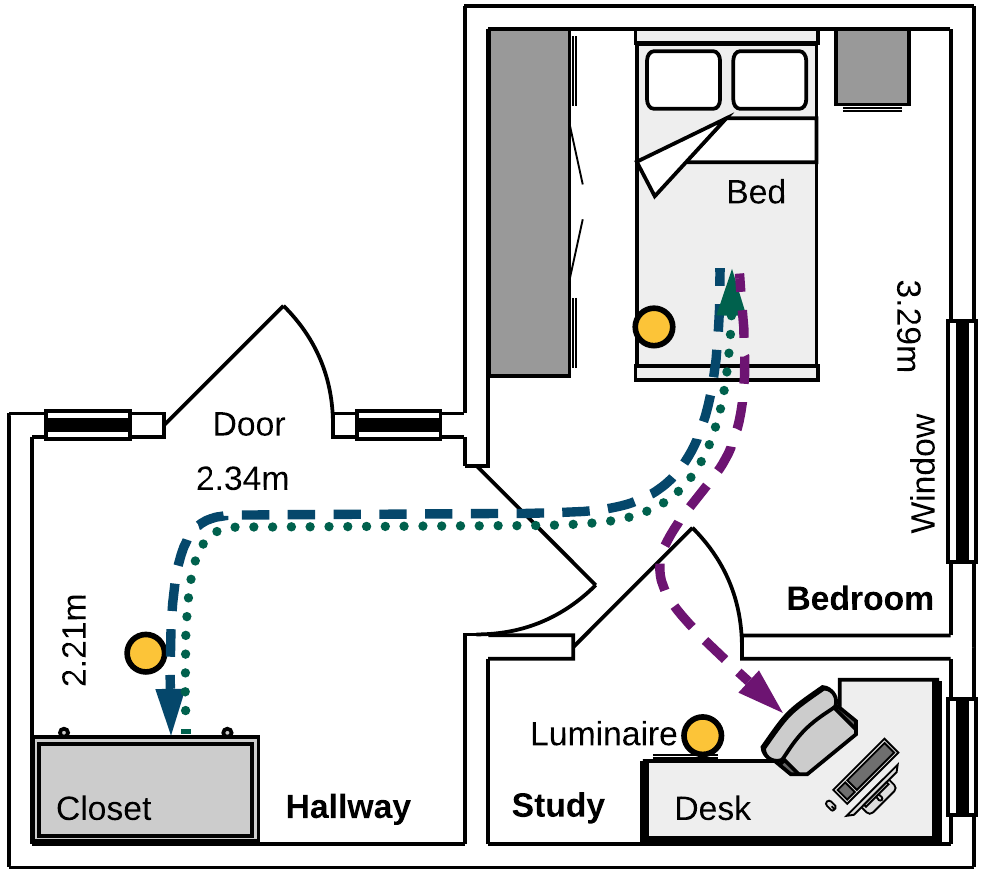}
    \caption{Floor plan of the dwelling used for experimentation. Evaluated scenarios were (a) walk from the bed to the closet; (b) walk from the closet to the bed; (c) walk from the bed to the desk chair.}
    \label{fig:floorplan}
\end{figure}

\subsection{Experimental setup}
The experiment was carried out in three rooms of an apartment located in a residential building. The layout of the relevant rooms is illustrated in Figure \ref{fig:floorplan}. For the sake of brevity, the kitchen and bathroom are not represented. Exterior and interior walls are constructed of brick and cement. Exterior walls have a thickness of 35cm, while the thickness of interior walls is 17cm. 

As shown in Table \ref{table:frequency_absorbtion}, building materials have an important deleterious effect on signal strength, making accurate localisation based on the raw values of signal strength difficult. As shown in Figure \ref{fig:floorplan}, one smart luminaire was ceiling-mounted in each room, replacing regular light bulbs. For the duration of the experiment, the luminaires had a stable Wi-Fi connection to the software server. 

The experiment consisted of three scenarios that were carried out by a healthy person using a smartphone as a BLE signal source. The three scenarios, also illustrated in Figure \ref{fig:floorplan} are: (a) start from the bed, walk to the closet and stop; (b) start from the closet, walk back to the bed and stop; (c) start from the bed, walk to the desk chair and stop. These were carried out in sequence, in this order, three times. Care was taken to ensure that every time the scenario started from the same initial conditions. Measurements were recorded every five seconds and the periods for each scenario were slightly different, depending on the distance the movement covered, as well as on how long the person was static at the end of the movement trail. The indoor locations are computed according to Formula \ref{eq:localisation} and using the Levenberg-Marquardt algorithm, by taking into consideration the last three recorded RSSI values by each luminaire, to attenuate single reading RSSI errors. Each scenario ends with the person staying in the same place for an interval between 50 to 70 seconds. This allows studying the effect movement has on the accuracy of localisation. It also allows evaluating how accuracy modifies when the subject ceases movement.

\begin{table}
\centering
\begin{tabular}{ |c|c| }
\hline
\textbf{Materials} & \textbf{Absorption Rate (dB)} \\
\hline
Brick wall & 6-15 \\
Cement wall & 4-6 \\
Glass wall & 6 \\
Metal door & 6-10 \\
Plasterboard & 3-5 \\
Window & 3   \\
\hline
\end{tabular}
\caption{Absorption of radio frequencies in construction materials relevant to our experiment \cite{buhagiar2018}.}
\label{table:frequency_absorbtion}
\end{table}

\subsection{Results}
\begin{figure}[h!]
\centering
\begin{subfigure}[b]{0.5\textwidth}
   \includegraphics[width=\linewidth]{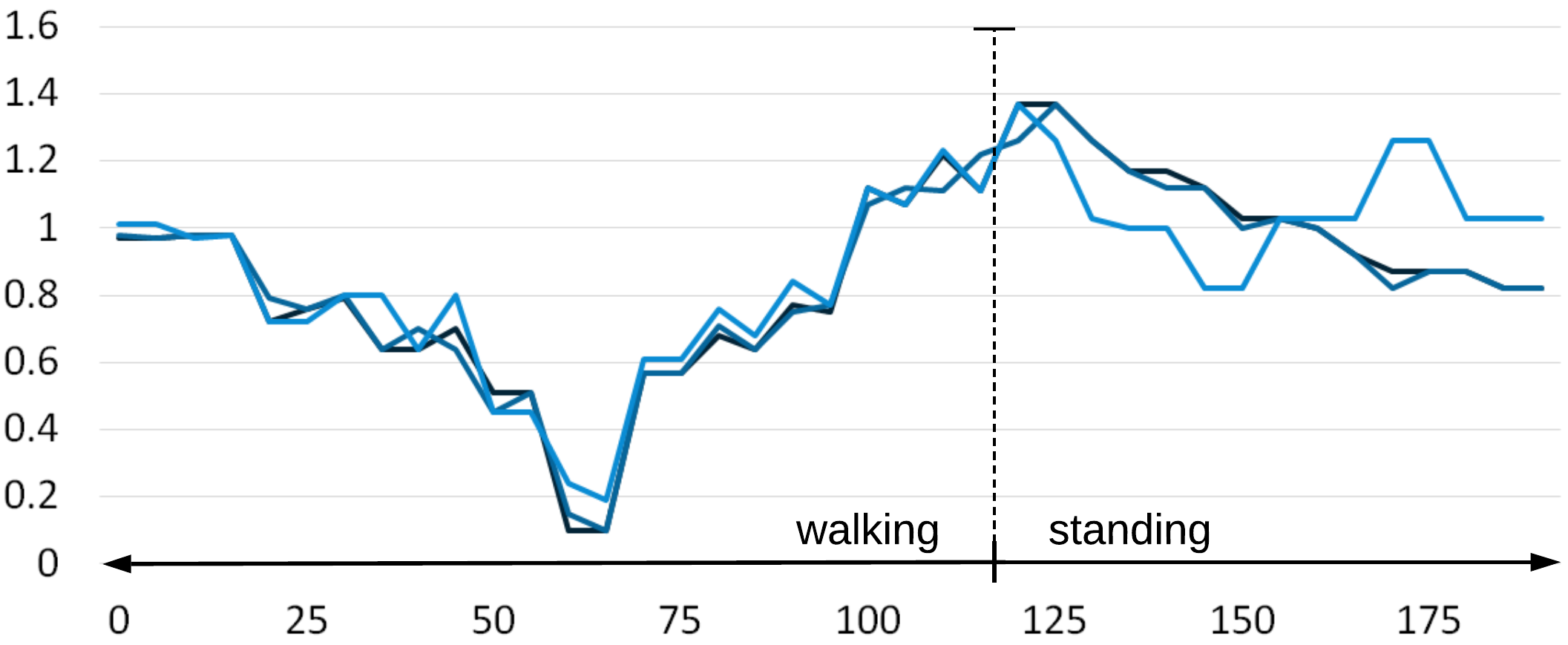}
   \caption{Person walks from the bed to the closet and stands in front of it.}
   \label{fig:interval1} 
\end{subfigure}
\par\medskip

\begin{subfigure}[b]{0.5\textwidth}
   \includegraphics[width=\linewidth]{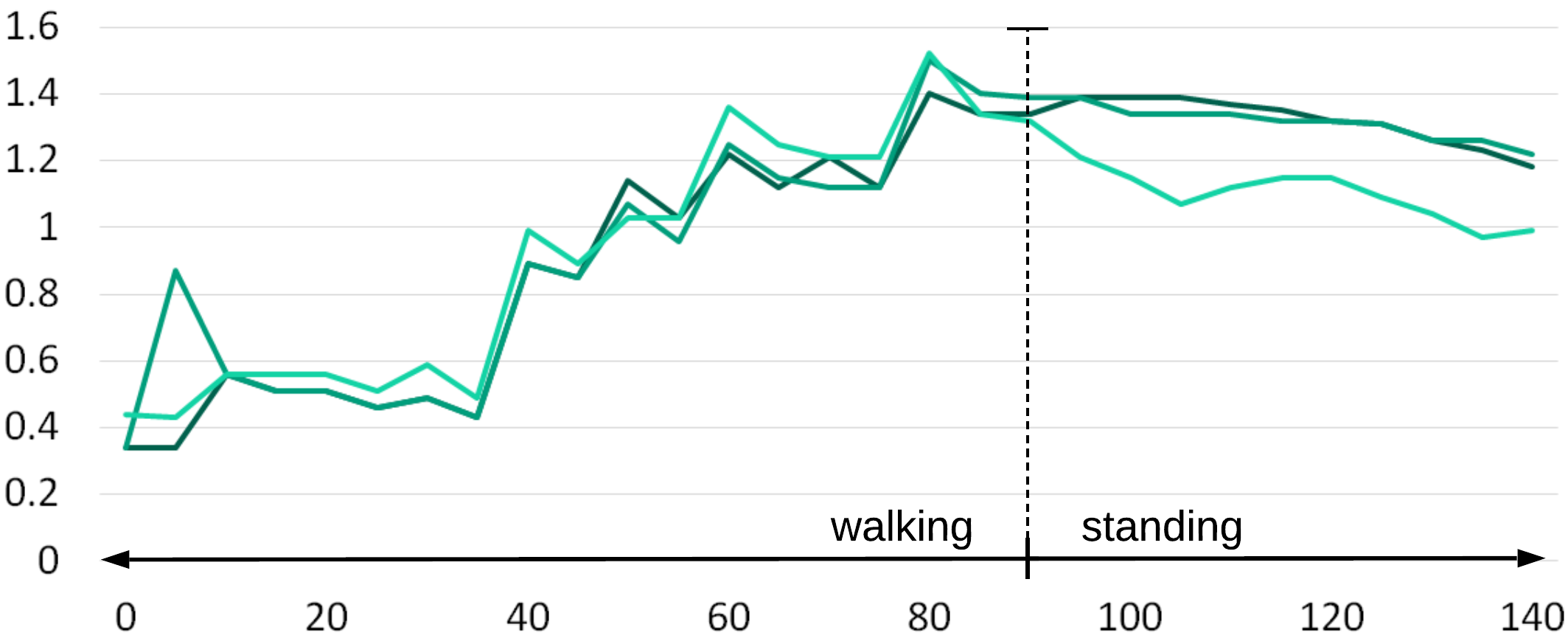}
   \caption{Person walks from the closet back to the bed and stands next to it.}
   \label{fig:interval2} 
\end{subfigure}
\par\medskip

\begin{subfigure}[b]{0.5\textwidth}
   \includegraphics[width=\linewidth]{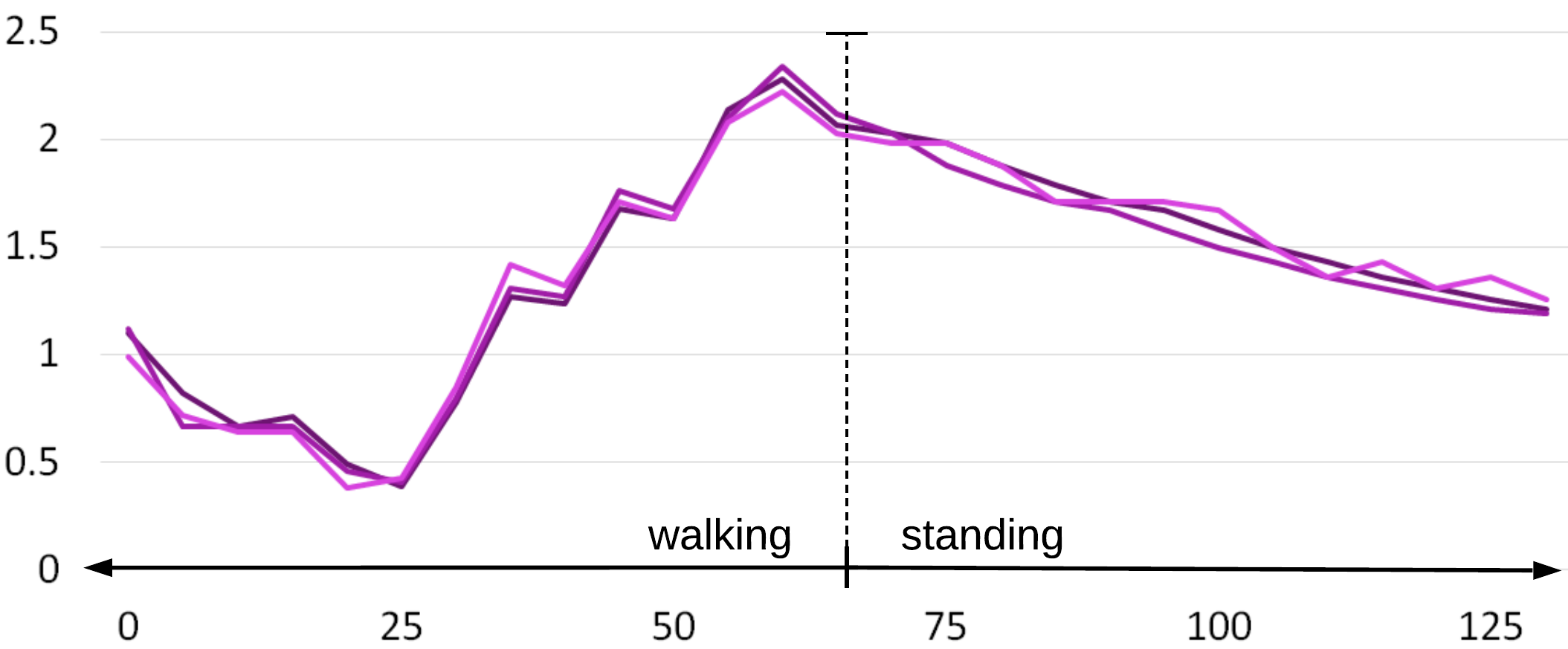}
   \caption{Person walks from the bed to the desk chair and sits down.}
   \label{fig:interval3} 
\end{subfigure}

\caption{The three scenarios of the experiment. Colours synchronised with Figure \ref{fig:floorplan}. The horizontal axis represents time, expressed in seconds and the vertical one is error, measured in meters. Each scenario was played out three times, as illustrated in the subfigures.}
\label{fig:errortimeline}
\end{figure}

Figure \ref{fig:errortimeline} illustrates the errors obtained in each of the three tested scenarios, considering three runs for each scenario. 

In all considered cases, computed indoor positions are coherent over repeated runs, namely the errors are highly consistent over time, which indicates the robustness of our method. This also leads us to believe that Bluetooth fingerprinting could be employed to further improve the accuracy of localisation during movement. We intend to further explore this in future research.

\begin{table}[h!]
\centering
\begin{tabular}{ |c|c|c|c|c| }
\hline
& \multicolumn{2}{c|}{Average error (m)} & \multicolumn{2}{c|}{Standard deviation (m)} \\
& Walking & Standing & Walking & Standing \\
\hline
1 & 0.76 & 1.05 & 0.27 & 0.17 \\
\hline
2 & 0.86 & 1.25 & 0.36 & 0.12 \\
\hline
3 & 1.17 & 1.61 & 0.60 & 0.28 \\
\hline
\end{tabular}
\caption{Average errors and standard deviations (over the three runs) obtained for all three scenarios.}
\label{table:errors}
\end{table}

Another important aspect targeted the impact of movement on localisation accuracy. Results obtained in all three cases indicate that the average error corresponding to the period the person was stationary are higher than when the person is moving. However, as depicted in Table \ref{table:errors} as well as Figure \ref{fig:errortimeline} plots, the standard deviation is lower. This indicates that the errors tend to stabilise and it can be noticed that they even slowly decrease during the time in which the target is immobile. As such, when measurements are recorded over longer periods of time they will tend to be more reliable and less prone to individual errors in measurement. The experiments show that without additional software processing \cite{marin2019indoor}, the system requires at least a minute to stabilise RSSI values, leading to more precise localisation results.

The differences in average errors between the moving and static experiments are particularly due to the higher errors obtained during movement, when the person was near doors, but which are also reflected in the error computation for the standing period. They are influenced by readings collected during movement, each position being computed by considering the past three RSSI values obtained from each of the three luminaires. These higher errors can also be observed in Figure \ref{fig:errortimeline} in each scenario, before the person stops. They are most likely due to signal absorption which occurs because walls around doors use iron-reinforced concrete pillars. While this interference happens during movement, it has a lasting impact and affects the calculations carried out when the person is already stationary.

\section{\uppercase{Conclusions}}
\label{sec:conclusion}
The presented work continues our initial evaluation regarding the accuracy of indoor localization of stationary subjects \cite{marin2019indoor}. Our first conclusion is that results obtained during the stationary phase remain similar to those obtained in previous work and provide room-level accuracy.

The second conclusion is that by itself, movement does not decrease localization accuracy. However, excessive multipath fading can be observed when the subject is close to architectural elements built from materials having high absorption rates. These cause a fluctuation of signal levels that persists across several readings, starting to stabilise once the subject is stationary. While our previous work \cite{marin2019indoor} shows that additional filtering can improve localization accuracy, these approaches must be further adapted and evaluated in the context of a moving subject. 

Future work will be carried out towards improving the accuracy of localization via filtering techniques adapted to subject motion together with evaluating the cost-benefit of location fingerprinting. Deep learning is an additional approach that can also be used to improve localisation results in an indoor noisy environment. Some experiments in this regard have already been performed \cite{marin2019indoor} for static targets, but we plan to extend the evaluation and to also consider data recorded during movement.

\vfill
\section*{\uppercase{Acknowledgements}}
\noindent This work was supported by a grant of the Romanian National Authority for Scientific Research and Innovation, CCCDI UEFISCDI, project number 46E/2015, \emph{i-Light - A pervasive home monitoring system based on intelligent luminaires}.

\bibliographystyle{apalike}
{\small
\bibliography{bibliography}}

\end{document}